\RequirePackage{fix-cm} 
\documentclass[a4paper, twoside, reqno, 12pt, dvips]{amsart}

\usepackage{fixltx2e}     

\usepackage{amssymb}
\usepackage{amsmath}
\usepackage{mathrsfs}

\usepackage{a4wide}

\usepackage{indentfirst}
\usepackage{graphicx}
\usepackage{psfrag}

\usepackage[usenames,dvipsnames]{pstricks}
\usepackage{epsfig}
\usepackage{pst-grad} 
\usepackage{pst-plot} 

\usepackage{subfigure}

\usepackage{longtable}

\usepackage[english]{babel}
\usepackage[latin1]{inputenc}

\usepackage{color}
\definecolor{oneblue}{rgb}{0,0.0,0.75}
\usepackage[colorlinks,
            urlcolor=oneblue,
            linkcolor=oneblue,
            citecolor=oneblue,
            bookmarksopen=false,
            pagebackref]{hyperref}

\numberwithin{equation}{section}

\newtheorem{remark}{Remark}

\newcommand{\um}{\bar{\/u}}

\newcommand{\vb}{\check{v}}

\newcommand{\uD}{\mathrm{D}}

\newcommand{\ud}{\mathrm{d}}

\renewcommand{\H}{\mathcal{H}}

\newcommand{\nus}{\tilde{\nu}}

\newcommand{\nub}{\check{\nu}}
\newcommand{\phim}{\bar{\phi}}
\newcommand{\x}{\boldsymbol{x}}                     
\newcommand{\phis}{\tilde{\phi}}
\newcommand{\phib}{\check{\phi}}
\renewcommand{\u}{\boldsymbol{u}}                   

\newcommand{\vmu}{\boldsymbol{\mu}}

\renewcommand{\epsilon}{\varepsilon}
\newcommand{\omb}{\boldsymbol{\omega}}
\newcommand{\scal}{\boldsymbol{\cdot}}              
\newcommand{\nab}{\boldsymbol{\nabla}}              
\newcommand{\vum}{\bar{\boldsymbol{\/u}}}
\newcommand{\half}{{\textstyle{1\over2}}}

\newcommand{\vmus}{\tilde{\boldsymbol \mu}}
\newcommand{\vmub}{\check{\boldsymbol \mu}}

\newcommand{\pd}[2]{\frac{\partial #1}{\partial #2}}

\begin{document}

\title[Improved Nonlinear Shallow Water Equations]{Shallow water equations for large bathymetry variations}

\author[D. Dutykh]{Denys Dutykh$^*$}
\address{LAMA, UMR 5127 CNRS, Universit\'e de Savoie, Campus Scientifique, 73376 Le 
Bourget-du-Lac Cedex, France}
\email{Denys.Dutykh@univ-savoie.fr}
\urladdr{http://www.lama.univ-savoie.fr/~dutykh/}
\thanks{$^*$ Corresponding author}

\author[D. Clamond]{Didier Clamond}
\address{Laboratoire J.-A. Dieudonn\'e, Universit\'e de Nice -- Sophia Antipolis, 
Parc Valrose, 06108 Nice cedex 2, France}
\email{diderc@unice.fr}
\urladdr{http://math.unice.fr/~didierc/}

\begin{abstract}
In this study, we propose an improved version of the nonlinear shallow water (or Saint-Venant) 
equations. This new model is designed to take into account the effects resulting from the large 
spatial and/or temporal variations of the seabed. The model is derived from a variational principle 
by choosing the appropriate shallow water ansatz and imposing suitable constraints. Thus, the 
derivation procedure does not explicitly involve any small parameter.
\end{abstract}


\maketitle

\section{Introduction}

The celebrated classical nonlinear shallow water (Saint-Venant) equations were derived for 
the first time in 1871 by A.J.C. de Saint-Venant \cite{SV1871}, an engineer working at Ecole 
Nationale des Ponts et Chauss\'ees. Currently, these equations are widely used in practice 
and the literature counts many thousands of publications devoted to the applications, 
validations or numerical solutions of these equations \cite{Synolakis1987, Zhou2002, 
Alcrudo2005, Dutykh2009a}.

Some important attempts have been also made to improve this model from physical point of view. 
The main attention was paid to various dispersive extensions of shallow water equations. The 
inclusion of dispersive effects resulted in a big family of the so-called Boussinesq-type equations 
\cite{Peregrine1967, Nwogu1993, Madsen03, Dutykh2007}. Many other families of dispersive wave 
equations have been proposed as well \cite{Serre1953, Green1976, Miles1985, Degasperis1999}.

However, there are a few studies which attempt to include the bottom curvature effect into the 
classical Saint-Venant or Savage-Hutter\footnote{The Savage-Hutter equations are usually posed 
on inclined planes and they are used to model various gravity driven currents such as snow 
avalanches.} \cite{Savage1989} equations. One of the first study in this direction is perhaps 
due to Dressler \cite{Dressler1978}. Much later, this research was pursued almost in the same 
time by Keller, Bouchut and their collaborators \cite{Keller2003,Bouchut2003}. We note that 
all these authors used some variant of the asymptotic expansion method. The present study is 
a further attempt to improve the classical Saint-Venant equations by including a better 
representation of the bottom shape. Moreover, as a derivation procedure we choose a variational 
approach based on the relaxed Lagrangian principle \cite{Clamond2009}.

In the next Section, we present the derivation and discussion of some properties of the improved 
Saint-Venant equations. Then we underline some main conclusions of this study in Section \ref{sec:concl}.

\section{Improved Saint-Venant equations derivation}

Consider an ideal incompressible fluid of constant density $\rho$. The horizontal independent 
variables are denoted by $\x = (x_1,x_2)$ and the upward vertical one by $y$. The origin of 
the cartesian coordinate system is chosen such that the surface $y=0$ corresponds to the still 
water level. The fluid is bounded below by a bottom at $y=-d(\x,t)$ and above by a free surface 
at $y=\eta(\x,t)$. Usually, we assume that the total depth $h(\x,t)=d(\x,t)+\eta(\x,t)$ remains 
positive $h(\x,t)\geqslant h_0 > 0$ for all times $t$. Traditionally in water wave modeling, the 
assumption of flow irrotationality is also adopted. The assumptions of fluid incompressibility 
and flow irrotationality lead to the Laplace equation for a velocity potential $\phi(\x,y,t)$.

It is well-known that the water wave problem possesses several variational structures 
\cite{Petrov1964,Luke1967,Zakharov1968}. Recently we proposed a relaxed Lagrangian variational 
principle which allows much more freedom in constructing approximations in comparison with 
classical formulations. Namely, the water wave equations can be derived by minimizing the 
functional $\iiint\mathscr{L}\,\ud^2\x\,\ud\/t$ involving the Lagrangian density \cite{Clamond2009}:
\begin{align}\label{defL}
\mathscr{L}\ =&\ (\eta_t+\vmus\scal\nab\eta-\nus)\,\phis\
+\ (d_t+\vmub\scal\nab d+\nub)\,\phib\ -\ \half\,g\,\eta^2\  \nonumber \\
&+\ \int_{-d}^{\,\eta}
\left[\,\vmu\scal\u-{\half}\/\u^2\,+\,\nu\/v-\half\/v^2\, +\,(\nab\scal\vmu+\nu_y)\,\phi\,\right]\ud\/y,
\end{align}
where $g$ is the acceleration due to gravity force and $\nab = (\partial_{x_1}, \partial_{x_2})$ 
denotes the gradient operator in horizontal Cartesian coordinates. Other variables \{$\u, v, 
\vmu, \nu$\} are the horizontal velocity, the vertical velocity and the associated Lagrange 
multipliers, respectively. The last two additional variables \{$\vmu,\nu$\} are called the 
pseudo-velocities. They formally arise as Lagrange multipliers associated to the constraints 
$\u=\nab\phi$, $v=\phi_y$. However, once these variables are introduced, the ansatz can be 
chosen regardless to their initial definition. The over `tildes' and `wedges' denote, respectively, 
a quantity traces computed at the free surface $y=\eta(\x,t)$ and at the bottom $y=-d(\x,t)$. We 
shall also denote below with `bars' the quantities averaged over the water depth. Note that the 
efficiency of the relaxed variational principle \eqref{defL} relies on the extra freedom for 
constructing approximations. 

In order to simplify the full water wave problem we choose some approximate, but physically 
relevant, representations of all variables. In this study, we consider very long waves in 
shallow water. This means that the flow is mainly columnar (Miles \& Salmon 1985) and that 
the dispersive effects are negligible. In other words, a vertical slice of the fluid moves 
like a rigid body.  Thus, we choose a simple shallow water ansatz, which is independent of 
the vertical coordinate $y$, and such that the vertical velocity $v$ equals the one of the 
bottom, i.e.,
\begin{equation}\label{anssha}
\phi\ \approx\ \phim(\x,t), \qquad \u\ =\ \vmu\ \approx\ \vum(\x,t), \qquad v\ =\ \nu\ \approx\ \vb (\x,t),
\end{equation}
where $\vb(\x,t)$ is the vertical velocity at the bottom. In the above ansatz, we take for 
simplicity the pseudo-velocity to be equal to the velocity field $\vmu=\u$, $\nu=v$. Note 
that in other situations they can differ (see \cite{Clamond2009} for more examples). With 
this ansatz, the Lagrangian density \eqref{defL} becomes
\begin{align}\label{defLsw}
\mathscr{L}\ =&\ (h_t+\vum\scal\nab h+h\nab\scal\vum)\,\phim\ -\ \half\,g\,\eta^2\ +\ 
\half\,h\,(\vum^2+\vb^2),
\end{align}
where we make appear the total water depth $h = \eta + d$.

\begin{remark}
Note that for ansatz \eqref{anssha} the horizontal vorticity $\omb$ and the vertical one 
$\zeta$ are given by:
\begin{equation*}
  \omb = \Bigl(\pd{\vb}{x_2}, -\pd{\vb}{x_1}\Bigr), \qquad
  \zeta = \pd{\um_2}{x_1} - \pd{\um_1}{x_2}.
\end{equation*}
Consequently the flow is not exactly irrotational in general. It will be confirmed below one more 
time when we establish the connection between $\vum$ and $\nab\phim$.
\end{remark}

Now we are going to impose one constraint by choosing a particular representation of the fluid 
vertical velocity $\vb(\x,t)$ at the bottom. Namely, we require fluid particles to follow the 
bottom profile:
\begin{equation}\label{eq:bot}
  \vb\ =\ -d_t\,-\,\vum\scal\nab d.
\end{equation}
This last identity is nothing else but the bottom impermeability condition within ansatz 
\eqref{anssha}. Substituting the relation \eqref{eq:bot} into Lagrangian density \eqref{defLsw}, 
the Euler--Lagrange equations yield
\begin{align}
\delta\/\phim\,:\quad 0\ &=\ h_t\ +\ \nab\scal[\,h\,\vum\,], \label{eqdphi}\\
\delta\/\vum\,:\quad \boldsymbol{0}\ &=\ \vum\ -\ \nab\phim\ -\ \vb\,\nab d, \label{eqdvu} \\
\delta\/\eta\,:\quad 0\ &=\ \phim_t\ +\ g\,\eta\ +\ \vum\scal\nab\phim\ -\ 
\half\,(\vum^2+\vb^2). \label{eqdeta}
\end{align}
Taking the gradient of \eqref{eqdeta} and eliminating $\phim$ from \eqref{eqdvu} gives us 
this system of governing equations
\begin{align}
h_t\ +\ \nab\scal[\,h\,\vum\,]\ &=\ 0, \label{eq:mas}\\
\partial_t\,[\,\vum\,-\vb\,\nab d\,]\ +\ \nab\,[\,g\,\eta\ +\ \half\,\vum^2\ 
+\ \half\,\vb^2\ +\ \vb\,d_t\,
] &=\ 0, \label{eq:qdm}
\end{align}
together with the relations
\begin{equation*}
  \vum\ =\ \nab\phim\ +\ \vb\,\nab d, \qquad
  \vb\ =\ -\,d_t\ -\ \vum\scal\nab d\ =\  
  -\frac{d_t\, +\, \nab\phim\scal\nab d}{1\, +\, |\nab d|^2}.
\end{equation*}

\begin{remark}
The classical nonlinear shallow water or Saint-Venant equations can be recovered substituting 
$\vb = 0$ into the last system, yielding
\begin{align*}
h_t\ +\ \nab\scal[\,h\,\vum\,]\ &=\ 0, \\
\vum_t\ +\ \nab\,[\,g\,\eta\ +\ \half\,\vum^2] &=\ 0.
\end{align*}
The last equation is most often written in the literature into the momentum flux form:
\begin{equation*}
\partial_t\,[\,h\,\vum\,]\ +\ \nab\,[\,h\,\vum^2\, +\, \half\,g\,h^2\,]\ =\ g\,h\,\nab d.
\end{equation*}
\end{remark}

\section{Secondary equations}

From the governing equations \eqref{eq:mas} and \eqref{eq:qdm}, one can derive an equation 
for the horizontal velocity $\vum$ and for the momentum density $h\vum$
\begin{equation}\label{eq:qdmhor}
  \vum_t\ +\ \vum\scal\nab\vum\ +\ g\,\nab\eta\ =\ \gamma\,\nab d \ +\ 
\vum\wedge(\nab\vb\wedge\nab d),
\end{equation}
\begin{equation}\label{eq:qdmflux}
\partial_t\,[\,h\,\vum\,]\ +\ \nab\,[\,h\,\vum^2\, +\, \half\,g\,h^2\,]\ =\ 
(g+\gamma)\,h\,\nab d\ +\ h\vum\wedge(\nab\vb\wedge\nab d),
\end{equation}
where $\gamma$ is the vertical acceleration at the bottom defined as
\begin{equation}\label{eq:acc}
  \gamma\ \equiv\ \frac{\uD\,\vb}{\uD\/t}\ =\ \vb_t\ +\ (\vum\scal\nab)\vb.
\end{equation}

\begin{remark}
In the right hand sides of \eqref{eq:qdmhor} and \eqref{eq:qdmflux}, the last term cancel out 
in one horizontal dimension. It can be easily seen from the following analytical representation 
which degenerates to zero in one horizontal dimension
\begin{equation*}
\vum\wedge(\nab\vb\wedge\nab d)\ \equiv\ (\nab\vb)\,(\vum\scal\nab d)\ 
  -\  (\nab d)\,(\vum\scal\nab\vb).
\end{equation*}
This property has the geometrical interpretation that $\vum\wedge(\nab\vb\wedge\nab d)$ is a 
horizontal vector orthogonal to $\vum$, thus vanishing for two-dimensional flows.
\end{remark}

One can also derive an equation for the energy flux
\begin{equation}\label{eq:energy}
\partial_t\left[\,h\,\frac{\vum^2 + \vb^2}{2}\,+\,g\,\frac{\eta^2-d^2}{2}\,\right]\, +\ 
\nab\scal\left[\left(\frac{\vum^2+\vb^2}{2} + g\eta\right)\,h\,\vum\,\right]\, =\ -\/(g+\gamma)\,h\, d_t.
\end{equation}
Obviously, the source term on the right-hand side vanishes if the bottom is fixed $d = d(x)$, 
or equivalently if $d_t = 0$. This last conservation law is closely related to the Hamiltonian 
structure of the improved Saint-Venant equations \eqref{eqdphi} -- \eqref{eqdeta}. Namely, 
these equations possess a canonical Hamiltonian structure for the variables $h$ and $\phim$, 
i.e.,
\begin{equation*}
  \pd{\,h}{\/t}\ =\ \frac{\delta\,\H}{\delta\/\phim}, \qquad
  \pd{\,\phim}{\/t}\ =\ -\/\frac{\delta\,\H}{\delta\/h},
\end{equation*}
where the Hamiltonian is
\begin{equation}\label{eq:Hamilt}
\H\ =\ \half\int\Bigl\{g(h-d)^2\, +\, h\,|\/\nab\phim\/|^2\,-\,
\frac{h\,[\,d_t\,+\,\nab\phim\scal\nab d\,]^2}{1\,+\,|\nab d|^2} \Bigr\}\,\ud^2\x.
\end{equation}
One can easily check, after computing the variations, that the Hamiltonian \eqref{eq:Hamilt} 
yields the equations
\begin{align*}
h_t\ &=\ -\nab\cdot\Bigl[\,h\nab\phim\ -\ \frac{d_t+\nab\phim\scal\nab d}
{1\,+\,|\nab d|^2}h\nab d\Bigr], \\
\phim_t\ &=\ -g(h-d)\ -\ \half\,|\nab\phim|^2\ +\ 
\frac{[\,d_t\,+\,\nab\phim\scal\nab d\,]^2}{1\,+\,|\nab d|^2},
\end{align*}
which are equivalent to the system \eqref{eqdphi}--\eqref{eqdeta}.

\begin{remark}
Rewriting the Hamiltonian \eqref{eq:Hamilt} in the equivalent form
\begin{equation}\label{eq:Hpos}
\H\ =\ \half\int\bigl\{\,g\eta^2\,+\,h\,\vum^2\,+\,h\,(\vb+d_t)^2\,-\,h\,d_t^{\,2}\,\bigr\}\,\ud^2\x,
\end{equation}
one can see that \eqref{eq:Hamilt} is actually positive definite if the bottom is static, i.e., 
if $d=d(x)$ or $d_t=0$. In other words, if there is no external input of energy into the system.
\end{remark}

\section{Hyperbolic structure}

For simplicity we consider in this section two-dimensional flows only, i.e., we have one 
horizontal dimension, say the $x_1$ and $u = u_1$ for brevity. Introducing the potential 
velocity $U=\phim_x$, we have
\begin{equation*}
\um\ =\ \frac{U\,-\,d_x\,d_t}{1\,+\,d_x^{\,2}}, \qquad 
\vb\ =\ -\,\frac{d_t\,+\,U\,d_x}{1\,+\,d_x^{\,2}}, \qquad
\um^2\, +\, \vb^2\ =\ \frac{U^2\,+\,d_t^{\,2}}{1\,+\,d_x^{\,2}},
\end{equation*}
and the equations of motion become
\begin{align}
\partial_t\,h\ +\ \partial_x\left[\,h\,\frac{U\,-\,d_x\,d_t}
{1\,+\,d_x^{\,2}}\,\right]\, &=\ 0, \\
\partial_t\,U\ +\ \partial_x\left[\,g\,(h-d)\, +\,\half\,U^2\ -\ \frac{(\,d_t\,+\,
U\,d_x\,)^2}{2\,+\,2\,d_x^{\,2}}\,\right]\, &=\ 0.
\end{align}

The Jacobian matrix of this quasilinear system along with its eigenvalues $\lambda_\pm$ 
can be easily computed:
\begin{equation}
\lambda_\pm\ =\ \um\ \pm\ c, \qquad c^{2}\ =\ g\,h\,[\,1\,+\,d_x^{\,2}\,]^{-1},
\end{equation}
where $c$ represents the gravity wave celerity. Note that for the classical Saint-Venant 
equations $c^{2}\ =\ g\,h$. Consequently, the gravity wave speed is slowed down in the new 
system by strong bottom variations.

\section{Conclusions}\label{sec:concl}

In this study, we derived a novel non-dispersive shallow water model which takes into account 
larger bathymetric variations. Previously, some attempts were already made to derive such systems 
for arbitrary slopes and curvature \cite{Bouchut2003, Keller2003}. However, our study contains 
a certain number of new elements with respect to the existing state of the art. Namely, our 
derivation procedure relies on a generalized variational structure of the water wave problem 
\cite{Clamond2009}. Moreover, we do not introduce any small parameter and our approximation is 
made through the choice of a suitable constrained ansatz which is partially equivalent to assume 
that the pressure is hydrostatic everywhere except at the bottom. Resulting governing equations 
have a simple form and hyperbolic structure. Another new element is the introduction of arbitrary 
bottom time variations. The Hamiltonian structure of the new model is provided as well. By 
comparing the vertical velocity profiles in our model and previous investigations 
\cite{Bouchut2003, Keller2003}, we conclude that a priori our model cannot be reduced to previous 
ones by choosing a particular asymptotic regime.

An interesting feature of the improved Saint-Venant equations is that this system is hyperbolic, 
like its classical counterpart. Thus, the same analytical and numerical methods can be used to 
study the new system. The wave propagation speed in the new system depends on the bottom gradient. 
This fact may have some important implications for practical problems and it will be investigated 
in the next study.

\section*{Acknowledgements}

D.~Dutykh acknowledges the support from French Agence Nationale de la Recherche, project 
MathOc\'ean (Grant ANR-08-BLAN-0301-01).

We would like to thank Professor Valeriy Liapidevskii for interesting discussions on gravity 
driven currents.

\bibliography{biblio}
\bibliographystyle{alpha}

\end{document}